\documentclass[sigconf]{acmart}

\usepackage{kotex}
\usepackage{enumitem}
\usepackage{multirow}
\usepackage{xcolor}

\newcommand{\HDY}[1]{{\color{black}#1}}

\newcommand{\ic}[1]{{\color{black}#1}}
\newcommand{\sy}[1]{{\color{black}#1}}

\AtBeginDocument{%
  }

\setcopyright{acmlicensed}
\copyrightyear{2018}
\acmYear{2018}
\acmDOI{XXXXXXX.XXXXXXX}
\acmConference[Conference acronym 'XX]{Make sure to enter the correct
  conference title from your rights confirmation email}{June 03--05,
  2018}{Woodstock, NY}
\acmISBN{978-1-4503-XXXX-X/2018/06}

\begin{document}

\copyrightyear{2025}
\acmYear{2025}
\setcopyright{acmlicensed}\acmConference[VRST '25]{31st ACM Symposium on Virtual Reality Software and Technology}{November 12--14, 2025}{Montreal, QC, Canada}
\acmBooktitle{31st ACM Symposium on Virtual Reality Software and Technology (VRST '25), November 12--14, 2025, Montreal, QC, Canada}
\acmDOI{10.1145/3756884.3766025}
\acmISBN{979-8-4007-2118-2/2025/11}
\title{Beyond the Portal:  Enhancing Recognition in Virtual Reality Through Multisensory Cues} 


\author{Siyeon Bak}
\email{123siyeon84@gmail.com}
\orcid{0009-0008-8845-2269}
\affiliation{%
  \institution{Hallym University}
  \city{Chuncheon}
  \country{Republic of Korea}
}

\author{Dongyun Han}
\email{dongyun.han@usu.edu}
\affiliation{%
  \institution{Utah State University}
  \city{Logan}
  \state{Utah}
  \country{United States}}

\author{Inho Jo}
\email{zjxps2007@gmail.com}
\affiliation{%
  \institution{Hallym University}
  \city{Chuncheon}
  \country{Republic of Korea}
}

\author{Sun-Jeong Kim}
\email{sunkim@hallym.ac.kr}
\affiliation{%
  \institution{Hallym University}
  \city{Chuncheon}
  \country{Republic of Korea}
}

\author{Isaac Cho}
\authornote{corresponding author}
\email{isaac.cho@usu.edu}
\affiliation{%
  \institution{Utah State University}
  \city{Logan}
  \state{Utah}
  \country{United States}}




\renewcommand{\shortauthors}{Trovato et al.}

\begin{abstract}

While Virtual Reality (VR) systems have become increasingly immersive, they still rely predominantly on visual input, which can constrain perceptual performance when visual information is limited. Incorporating additional sensory modalities, such as sound and scent, offers a promising strategy to enhance user experience and overcome these limitations. 
This paper investigates the contribution of auditory and olfactory cues in supporting perception within the portal metaphor, a VR technique that reveals remote environments through narrow, visually constrained transitions. 
We conducted a user study in which participants identified target scenes by selecting the correct portal among alternatives under varying sensory conditions.
The results demonstrate that integrating visual, auditory, and olfactory cues significantly improved both recognition accuracy and response time. 
These findings highlight the potential of multisensory integration to compensate for visual constraints in VR and emphasize the value of incorporating sound and scent to enhance perception, immersion, and interaction within future VR system designs.


\end{abstract}


\begin{teaserfigure}
  \centering
  \includegraphics[width=.95\textwidth]{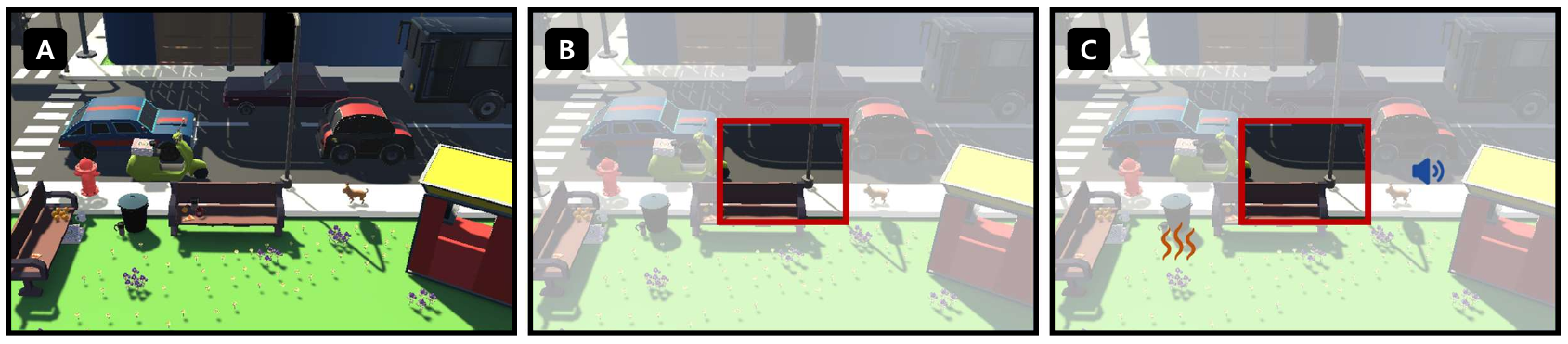}
  \vspace{-0.3cm}
  \caption{Experimental environment in the multisensory portal study. (A) The city street scene contained contextual elements such as cars, a flower, and a dog. (B) Through the portal, participants viewed only a restricted subset of the scene. (C) In multisensory conditions, additional cues could be introduced (e.g., auditory barking from the dog or a floral fragrance from the flower) to support scene identification under limited visual information.}
  \Description{Portals can provide multisensory experiences. They can also not provide multisensory experiences.}
  \label{fig:teaser}
\end{teaserfigure}


\maketitle

\section{Introduction}

The portal metaphor in Virtual Reality (VR) presents a powerful technique for interaction and navigation by visually connecting spatially separated virtual regions. Through portals, users can observe and even traverse distant locations, enabling efficient access to remote scenes without the need for continuous physical or virtual travel~\cite{han2022portal, ablett2023point}. This approach is particularly valuable in confined or complex VR environments where physical space is limited. However, portals are often constrained in size, offering only a narrow field of view (FoV) as shown in Figure ~\ref{fig:teaser}. Such spatial restrictions limit users’ ability to perceive and interpret the context of remote scenes before engaging with them directly, thereby reducing spatial awareness and increasing cognitive effort~\cite{Koltai2019Procedurally, Wu2018Efficient, Wang2019Occlusion, Wu2018Anchored}. 

To address these challenges and \ic{examine how cross-modal priming can support scene recognition when visual input is restricted}, we explore \textit{multisensory portals}, portals augmented with auditory and olfactory cues. Although most VR systems primarily rely on visual and auditory modalities, real-world perception is inherently multisensory. Human cognition is shaped by the integration of sensory inputs, including vision, hearing, and smell, which operate in conjunction to build a coherent understanding of the environment~\cite{Damon2021Olfaction}.

While many VR systems support visual and auditory channels, prior research has emphasized the need for additional sensory modalities to better simulate real-world experiences~\cite{fontaine1992experience, witmer1998measuring}. In particular, olfactory cues have demonstrated unique benefits: they capture user attention~\cite{Yang2024Olfaction}, evoke affective and contextual memory~\cite{Low2013Olfactive}, and enhance autobiographical memory retrieval when combined with visual stimuli~\cite{Sasu2024Enhancing}. Similarly, auditory cues provide temporal and spatial signals that help direct user attention and support environmental understanding~\cite{ranasinghe2018season, ebrahimi2016empirical, hagelsteen2017faster}.

This work investigates whether augmenting portals with olfactory and auditory cues can enhance users’ perception of scenes beyond the portal in VR \ic{through cross-modal priming}, particularly under conditions of constrained visual information. While prior research has largely employed multisensory integration to improve realism and immersion~\cite{ranasinghe2018season, ebrahimi2016empirical, hagelsteen2017faster}, we reframe the portal as a perceptual bottleneck and examine whether sensory augmentation can expand its informational utility. We conducted a formal user study to evaluate how the integration of sensory cues influences spatial recognition and object identification through portals. 
In our study, participants selected the correct portal from four alternatives based on a brief textual description of a remote scene. The results show that participants more accurately and efficiently identified target scenes when portals were enhanced with multisensory cues, demonstrating the potential of sensory augmentation to mitigate visual limitations and improve perceptual performance in constrained VR interfaces.

\section{Related Work}

\subsection{MultiSensory Cues in VR}

Although many VR systems still focus on providing users with an immersive experience through visual and auditory stimuli, growing studies highlight the importance of incorporating additional senses like olfaction and haptics in VR experience as multisensory cues can enhance user performance in VR tasks ~\cite{stein2008multisensory, petrini2012vision, arnold2018you}.
Moreover, multisensory cues enhance users' experiences by fostering a greater sense of naturalness, immersion, and realism~\cite{ranasinghe2018season, ebrahimi2016empirical, hagelsteen2017faster}. Previous research has shown that increasing the variety of sensory stimuli contributes to a more realistic virtual experience, which in turn leads to significantly higher user satisfaction~\cite{apostolou2022systematic, melo2020multisensory}.

While many earlier works explore tactile cues,  olfactory cues have also been explored to enhance the sense of immersion and the perceived presence of virtual objects in VR~\cite{munyan2016olfactory, persky2020olfactory}. 
Olfactory cues also play a crucial role in enhancing autobiographical memory recall and emotional responses \cite{Sasu2024Enhancing}, and have also been shown to improve cognitive retention during language learning \cite{xia2024amplifying}. Given these benefits, previous studies have proposed VR applications that incorporate olfactory cues for the treatment of post-traumatic stress disorder (PTSD)~\cite{aiken2015posttraumatic} and cognitive therapy for individuals with autism spectrum disorder (ASD)~\cite{barros2020giving}.
Additionally, the combined use of olfactory and tactile cues has been found to effectively elicit specific emotional responses, such as trust or disgust~\cite{alshaer2025role}.


Previous studies have shown that multisensory cues help improve users' task performance, and that smell in particular aids memory recall and emotional responses~\cite{chanes2016redefining, sullivan2015olfactory, amores2017essence, holland2005smells}. 
There have also been studies showing that smell influences user behavior~\cite{quintana2019effect}.  Our study uses olfactory cues as one of the multisensory cues and focuses on whether participants can recognize a space or object and select the corresponding button. This is similar to the study by Persky and Dolwick \cite{persky2020olfactory}, which found that olfactory cues influence participants' food choices, but our study differs in that we provide olfactory cues to help participants select the corresponding object.


\subsection{Olfactory Device for VR}


Researchers have developed various olfactory devices that release scents through methods such as vapor diffusion~\cite{ranasinghe2018season, javerliat2022nebula, myung2023enhancing}, as well as by heating solid~\cite{dobbelstein2017inscent} or liquid materials~\cite{covington2018development}.
The olfactory design space includes four key aspects: chemical, emotional, spatial, and temporal~\cite{maggioni2020smell}, which correspond to scent types, users’ emotional reactions, the spatial origin of the scent, and how long the scent is perceived. 
In everyday environments, these aspects help individuals navigate spaces and interpret scent-related information such as type, strength, and blends~\cite{amores2017essence, dobbelstein2017inscent}. 
Recreating such olfactory dynamics in VR, however, poses significant challenges due to technological limitations, particularly in managing scent intensity and offering a broad scent range. Nonetheless, recognizing the value of olfactory input is essential for improving the quality of VR experiences~\cite{covarrubias2015vr, tsai2021does}.
In this study, we designed an olfactory device that is attachable to the bottom of the head-mounted display (HMD) based on a previous approach ~\cite{myung2023enhancing}. 
This allows scents to be delivered instantly and ventilated within the VR environment.


\subsection{Portal Interaction in VR}
In VR, users typically navigate virtual environments through physical movements, but these are limited by the real-world physical space. To address these constraints, a range of virtual locomotion techniques have been developed, including steering, teleportation, and portal \cite{cherni2020literature, huber2023exploring, atkins2021continuous}. Among them, portal facilitates navigation of a broader virtual space by connecting disparate virtual environments, even within confined spaces, thereby enabling users to experience transitioning between virtual environments while traversing a confined space \cite{atkins2021continuous}.

Research on portal movement is being conducted with a focus on operability, motion sickness prevention, and maintaining immersion. However, in complex environments or those with multiple users, portal-based movement can cause disorientation or confusion regarding spatial awareness \cite{Koltai2019Procedurally, thanyadit2020substituting}. In addressing this issue, various methods have been proposed and studied, including minimizing physical movement, limiting the FoV, or providing visual effects \cite{gulcu2022infinite, atkins2021continuous, thanyadit2020substituting, peng2020walkingvibe}.

The impact of portals on users is also subject to variation depending on the portal's dimensions. However, smaller portals may impose limitations on user immersion due to their constrained FoV, necessitating physical movement to verify the space beyond the portal \cite{Koltai2019Procedurally, gulcu2022infinite, ghosh2024merp}. Despite their comparatively restricted field of view, small-sized portals demonstrated a higher degree of efficacy in mitigating directional confusion when compared to their larger counterparts. Conversely, the use of larger-sized portals has been demonstrated to enhance immersion and improve spatial awareness \cite{atkins2021continuous, husung2019portals, cisternino2019virtual}. However, extant studies have indicated that users experience an increase in cognitive load due to the increased amount of information they must process concurrently \cite{ablett2023point}.
This paper addresses the challenge of the limited FoV inherent in small-sized portals by integrating auditory and olfactory stimuli into the navigation process to support object and location identification.


\section{User Study:  Multisensory Cues in Portals}

Our user study investigates whether augmenting a VR portal with auditory and olfactory cues enhances users’ understanding of a scene viewed through the portal. 
We compare a visual-only condition against multisensory conditions to explore the potential effectiveness of multimodal cues in improving the portal experience under limited FoV constraints. This study addresses the following research questions (RQs):
\begin{itemize}
\item[\textbf{RQ1}:] How do multisensory cues help users grasp the context of a scene that is only partially visible through the portal?

\item[\textbf{RQ2}:] What are the specific contributions of auditory and olfactory cues to portal-based interaction?
\end{itemize}

\subsection{Study Design}

Our study employs a within-subjects design based on four sensory cue conditions augmented in the portal. When a participant approaches within 1 meter of the portal, they can perceive additional auditory and/or olfactory cues in addition to the visual cue, depending on the assigned condition. The conditions are as follows:

\begin{itemize}[leftmargin=0.15in, noitemsep]
    \item Visual only \textbf{(V)}: This is the baseline condition where participants only see the visual scene inside each portal. No additional auditory or olfactory cues are provided. 
    \item Visual plus olfactory \textbf{(VO)}: In this condition, participants can see the scene inside the portal and perceive a scent corresponding to the scene when they are within 1 meter of the portal.   
    \item Visual plus auditory \textbf{(VA)}: In this condition, participants are presented with both visual and auditory cues. As they approach the portal, they hear sounds associated with the scene inside. 
    \item Visual plus auditory plus olfactory \textbf{(VAO)}: This is the fully multisensory condition, where participants receive visual, auditory, and olfactory cues simultaneously. 
\end{itemize}

In this study, we assumed that the portal presents a scene with a limited FoV, following the approach adopted in prior work~\cite{han2022portal, ablett2023point}. The portal was configured as a square measuring 1 meter by 1 meter—twice the size used in Ablett et al.'s design~\cite{ablett2023point} and comparable in scale to the circular portal used by Han et al.~\cite{han2022portal}. This dimension was selected to provide a constrained yet sufficiently detailed view, balancing visual limitation, essential for evaluating multisensory cue effects, while ensuring that the content within the portal remains both perceptible and semantically interpretable to participants.

\begin{figure*}
  \centering 
  \includegraphics[width=.95\linewidth]{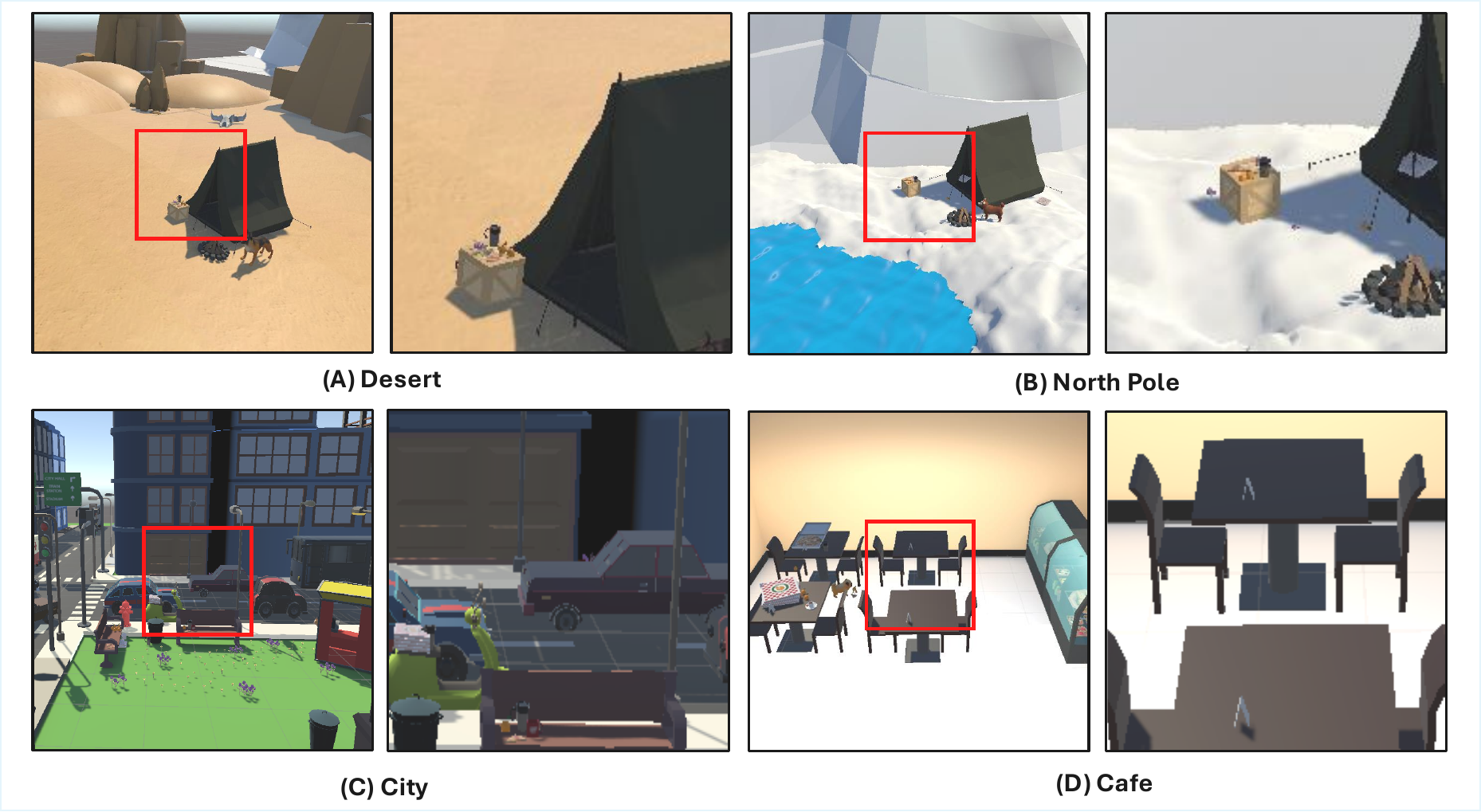}
  \caption{In the left images, the red box outlines the limited field of view available through the portal, which reveals only part of the scene. The right images show the user’s perspective inside the portal, where additional details may become visible as the user moves. Four distinct virtual scenes were used in the study: (A) Desert, (B) North Pole, (C) City, and (D) Cafe.
  }
  \label{fig:Experiment_scenes}
\end{figure*}

We designed four visually distinct virtual environments, Desert, North Pole, City, and Cafe, as illustrated in Figure~\ref{fig:Experiment_scenes}.  
They were selected for their easily recognizable visual characteristics. 
\HDY{Each environment could include various virtual objects.
Among them, some objects presented their corresponding auditory or olfactory cues, called olfactory and auditory elements in this work.
The auditory elements consisted of four types: a dog vocalization, a bird vocalization, a person coughing, and music, represented in the scene by a virtual dog, bird, human, and speaker, respectively.
The olfactory elements consisted of virtual objects such as coffee, orange, flowers, and pizza, with each contributing a distinct scent\footnote{All scent products were procured from \url{https://www.esfood.kr}.}. 
Participants were able to see virtual environments and objects through portals. 
Participants could move around the VR scene to view the portals from different angles and clearly identify these objects within the portals. This could be important in the sole V condition, where no auditory or olfactory cues were available.}


\begin{figure}[t]
  \centering 
  \includegraphics[width=.9\linewidth]{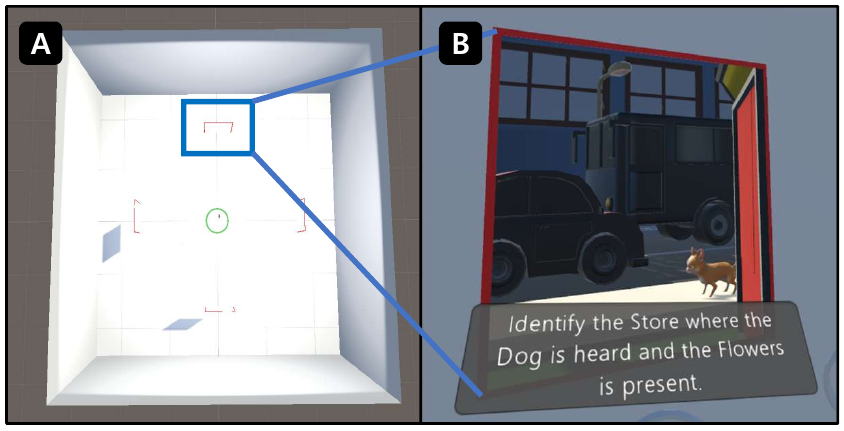}
  \caption{(A) The VR scene used in the formal user study, where four portals were positioned in the virtual room, each showing a different scene. (B) Example of the city scene displayed through one portal.
  }
  \label{fig:Experiment Scene1}
\end{figure}


\subsection{Task}



In each task, four portals were situated within a virtual room, as shown in Figure~\ref{fig:Experiment Scene1}A. Participants began exploring the portals from the center of the room (i.e., the green circle in Figure~\ref{fig:Experiment Scene1}A).
Two randomly selected environments out of the four were displayed through two portals each. 
Each auditory and olfactory element was placed in a separate portal along with various virtual objects.


Depending on the sensory condition, participants could perceive a scent and/or hear a sound when they were within 1 meter of a portal, or they might not receive any additional cues other than a visual cue. 
\HDY{For each trial, participants received the following prompt: ``Identify the [\textit{virtual  environment}] where the [\textit{object 1}] is heard and the [\textit{object 2}] is present.'' 
While both objects could be verified visually, \textit{object 1} was additionally represented through auditory cues and \textit{object 2} through olfactory cues, depending on the multisensory condition.}
They were instructed to select the portal that matched the description in the prompt, then either proceed to the next task or complete the study.

The experimental design aimed for simplicity and intuitiveness, with users selecting one of four portals. The decision was predicated on two considerations. 
First, the provision of the same sensory information to all four portals was implemented to prevent the mixing of types and to facilitate a clear understanding of the effects of sensory information. 
Secondly, the provision of information regarding various objects for each portal was intended to facilitate user comprehension of the inquiries and enable intuitive selections. 
The objective of this study was to ensure that the research findings were both clear and intuitive in terms of user perception, with the emphasis on sensory information.

\subsection{Measure and Hypothesis}

Quantitative measures are the participants' response accuracy for portal selection, task completion time, and confidence rating. 
Two subjective indicators are the NASA Task Load Index (NASA-TLX) \cite{hart1988development} and the Simulator Sickness Questionnaire (SSQ) \cite{kennedy1993simulator}.



\begin{itemize}[leftmargin=0.15in]
\setlength\itemsep{0pt}

\item \textbf{Accuracy Rate:} The proportion of correct responses out of 10 trials per sensory condition, reported as a percentage (0–100\%). Each trial required participants to select the correct portal based on a given cue, and accuracy reflects the effectiveness of cue interpretation.
\item \textbf{Task Completion Time:} The time (in seconds) taken by participants to complete each trial, measured from the onset of the task to the submission of a response. Each participant completed 10 trials per sensory condition, and the average completion time was computed per condition.
\item \textbf{Confidence Rating:} After each trial, participants rated their confidence in their choice using a 7-point Likert scale (1 = "Not confident at all", 7 = "Extremely confident"). These ratings were collected across 10 trials per sensory condition and averaged to evaluate perceived certainty.
\item \textbf{NASA-TLX:} A standardized subjective workload assessment tool comprising six subscales: mental demand, physical demand, temporal demand, performance, effort, and frustration. Participants completed the NASA-TLX after finishing all trials within each sensory condition to evaluate perceived workload.
\item \textbf{SSQ:} A 16-item validated instrument used to assess symptoms of simulator or motion sickness (i.e., nausea, oculomotor discomfort, disorientation) experienced during VR tasks. The SSQ was administered after completing all trials in each condition to monitor adverse effects related to VR exposure. 

\end{itemize}

After completing the task under each sensory condition, participants were asked to rate how helpful the sensory cues were using a 7-point Likert scale (1 = Not at all helpful, 7 = Extremely helpful). In addition, participants were also prompted to provide open-ended feedback describing how or why the cues were (or were not) helpful during the task. 
This experiment was designed to prove the following hypothesis.

\begin{enumerate}
[label=\textbf{H\arabic*.}, leftmargin=0.35in]
\setlength\itemsep{0pt}
\item  Participants are expected to have higher accuracy rates and faster completion times in multisensory conditions (VO, VA, and VAO) compared to the V condition. The presence of congruent auditory and/or olfactory cues is anticipated to provide additional contextual information, thereby facilitating quicker and more accurate scene recognition.

\item Participants will report higher confidence ratings in their responses under multisensory conditions compared to the visual-only condition. This is because supplementary sensory inputs are likely to reduce uncertainty and reinforce the correctness of participants’ interpretations, resulting in greater perceived confidence during decision-making.

\item Participants are expected to show lower mental and physical load in multisensory conditions (VO, VA, and VAO) than in the V condition. This is because they will recognize the location more quickly and feel less burden when other senses are provided than when only vision is provided to recognize the location.

\item The VAO condition was expected to yield the highest SSQ scores, with VA and VO producing similar but slightly higher scores than the V condition. This expectation is based on the assumption that the addition of auditory and olfactory stimuli may introduce mild sensory conflicts or increase perceptual load, potentially leading to elevated simulator sickness symptoms compared to the baseline condition.

\end{enumerate}

\subsection{Apparatus}



\begin{figure}[t]
  \centering
  \includegraphics[width=\linewidth]{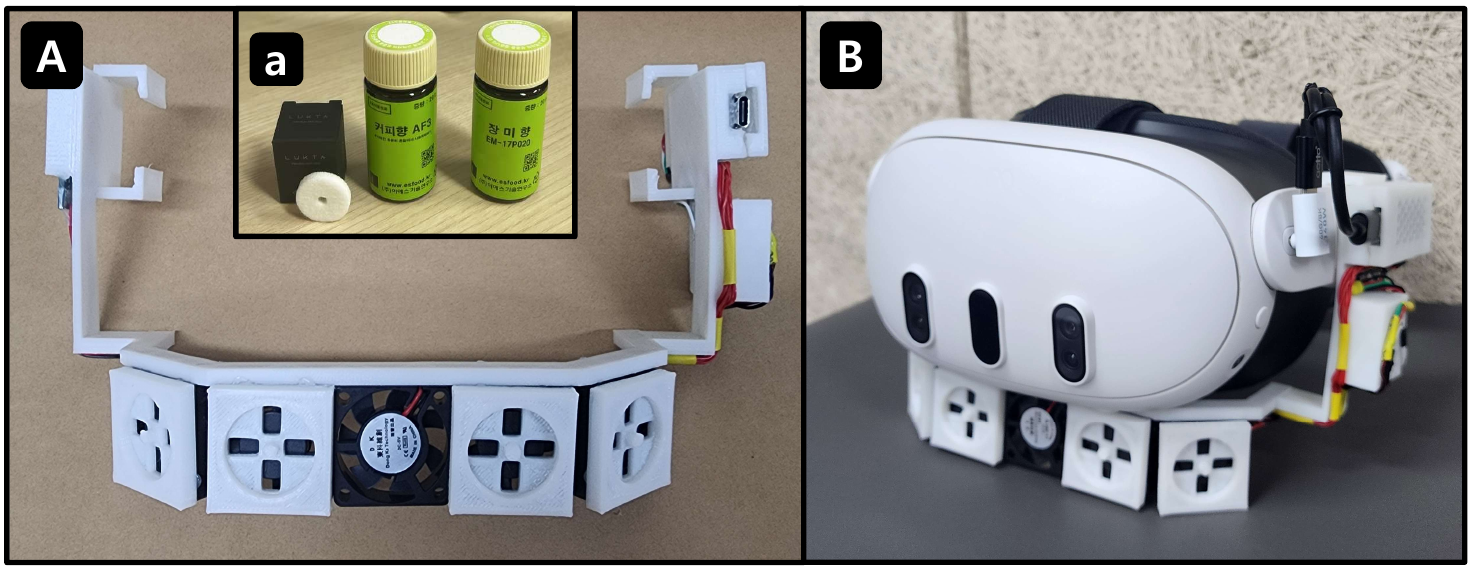}
  \caption{The olfactory device used in the study: (A) The device contains five fans, with the center fan used for ventilation and the other four for scent delivery. (a) Fragrance oils and cotton pads used for scent delivery. The cotton pads were soaked in fragrance oils and placed inside a 3D-printed cartridge positioned in front of the four scent-dispersing fans.
  (B) The device can be attached beneath the Meta Quest 3 headset.
  }\label{fig-olfactory-device}
  \vspace{-0.5cm}
\end{figure}

This study used a Meta Quest 3, a desktop to run the VR scene, and an olfactory device.
Meta Quest 3 has a  110$^\circ$ of horizontal FoV and a 96$^\circ$ of vertical FoV,  with a resolution of 2064 × 2208 pixels per eye.
The virtual scene for the study was developed in Unity version 2021.3.19f1 using a high-performance PC equipped with an AMD Ryzen 5 4600H CPU, 16GB RAM, and a GeForce RTX2060, and Windows 11 was installed as the operating system.

The olfactory device used in this study was adapted for integration with the Meta Quest 3, based on the design proposed by Myung et al. \cite{myung2023enhancing} (Figure~\ref{fig-olfactory-device}). 
It comprises two main components: the \textit{Scent Control Unit} and the \textit{Scent Delivery Unit}, which are attached to the side and lower sections of the Quest 3, respectively. 
It is capable of delivering up to four distinct scents and includes a built-in ventilation system to manage scent dispersion and clearance. 
The total weight of the device is approximately 102 grams.

\textit{The Scent Control Unit} includes an Arduino Nano~\cite{ArduinoNano}, a Bluetooth module (HC-06) \cite{bluetoothmodule}, five 5V fans, and four scent modules. The Arduino Nano features a compact size, lightweight design, and low power consumption. It is powered directly via a USB connection to Meta Quest 3, requiring no external power supply. 
Communication between the device and the VR application—for both scent emission and ventilation—is handled via Bluetooth serial communication through the Arduino Nano.

\textit{The Scent Delivery Unit} consists of five fans, four scent modules, and supporting frames as shown in Figure \ref{fig-olfactory-device}A.  Each fan measures 30mm \texttimes{} 30mm (width \texttimes{} height) and operates at 5V. Among the five fans, four are used for dispersing scents, while the central fan is dedicated to ventilation. Unlike the scent fans, the ventilation fan is installed in reverse to help clear residual odors and maintain air circulation. Each scent module contains a cotton pad soaked in scented oil (Figure~\ref{fig-olfactory-device}-a). When the system is activated, the fans generate airflow that disperses the scent into the environment, providing users with olfactory feedback.




\subsection{Participants}



Initially, 28 participants were recruited through university-affiliated social media channels. One participant was excluded from the analysis due to dizziness experienced during the task, resulting in withdrawal from the study. Consequently, data from 27 participants were included in the final analysis (21 males, 6 females; M = 23.04, SD = 2.89, age range = 18–27).
\sy{
This number of participants satisfied the requirements of our power analysis conducted with G*Power 3.1 for a within-subject ANOVA study. The parameters were: effect size f = 0.40, α error probability = .05, power = .80, number of groups = 4, number of measurements = 2, correlation among repeated measures = .50, and nonsphericity correction ε = 1. This analysis yielded a minimum sample size requirement of N = 24.
}
All participants reported normal vision, hearing, and sense of smell. Nineteen participants had prior experience with VR. Each participant received approximately \$7.27 for their participation, which reflected the local hourly minimum wage in the country where the study was conducted. 


\subsection{Study Procedures}

The study lasted approximately 60 minutes (IRB: HIRB-2024-038). Upon arrival, participants sign an informed consent form and complete a demographic questionnaire. An instructor then provides an overview of the study, including its objectives, procedures, scent types, and the devices used. To ensure participants were familiar with the scents, they were asked to smell the four scents that would be used in the task. 
Following that, participants engaged in a training session to practice viewing and selecting the portal and learning how to use the VR controllers to complete the task. 
The training session took about five minutes.
Before proceeding to the main session, participants completed the SSQ questionnaire to assess their baseline state. 


\begin{figure*}
  \centering
  \includegraphics[width=\linewidth]{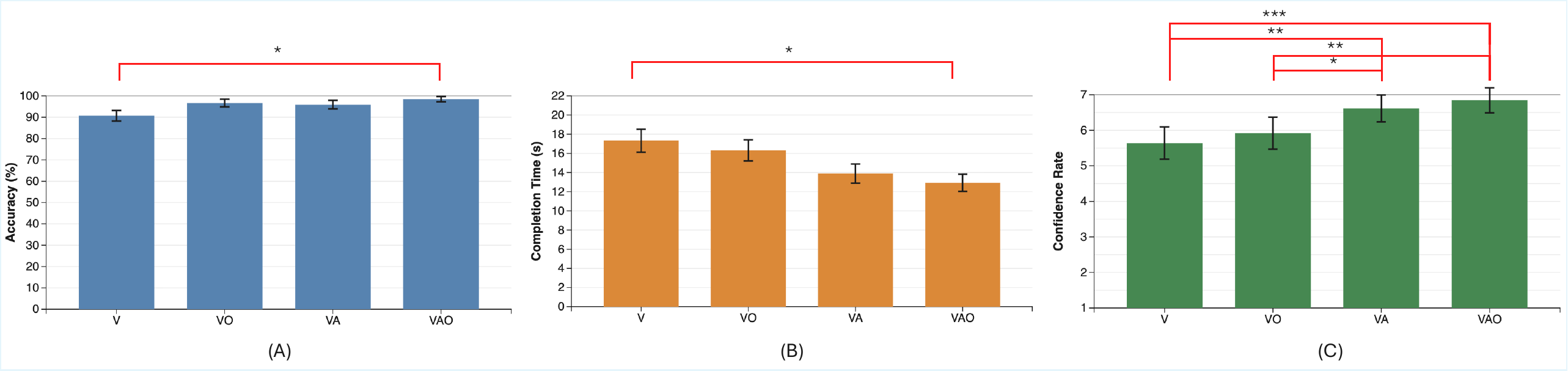}
 \caption{Study results across conditions. (A) Accuracy rate, (B) task completion time, and (C) confidence rating. Error bars represent the 95\% confidence interval. * indicates statistical significance between conditions ($^{***}p < .001$, $^{**}p < .01$, $^{*}p < .05$).}
 \label{Study1-Accuracy-Rate}
\end{figure*}

The main session consisted of four sub-sessions. It was conducted with participants standing. 
In each sub-session, participants performed ten tasks under one of the four sensory conditions. 
After completing each sub-session, they responded to the NASA-TLX and SSQ questionnaires. Then, they took a 3-minute break before proceeding to the next sub-session. During these breaks, the windows in the room where the experiment was conducted were opened for ventilation. And, the participants took off their HMD and took a break. 
After completing all tasks, participants fill out a post-questionnaire. 
At the end of the study, the instructor ventilates the room and sanitizes the olfactory device with alcohol to eliminate any residual scents.

\subsection{Analysis}

We report the results analyzed with a one-way repeated measures Analysis of Variance (ANOVA) test at the 5\% significance level. 
The degrees of freedom are corrected using the Greenhouse-Geisser correction to protect against violations of the sphericity assumption. 
A post-hoc test \sy{using the Bonferroni correction for multiple comparison is employed} when a significant effect was observed at the 5\% significance level. 
We report the mean, minimum, and maximum values for each condition.
\subsection{Result}


\paragraph{\textbf{Accuracy Rate}}

There was a main effect on the accuracy rate ($F(2.239, 58.213)$= 5.467, $\eta_p^2$=.174, $p$=.005). 
\sy{
The accuracy rate was lowest in V (M = 90.741 [86.502, 94.980]), followed by VA (M = 95.926 [92.970, 98.882]), VO(M = 96.667 [93.197, 100.136]), and highest in VAO(M = 98.519 [97.086, 99.951]).
Pairwise comparisons revealed that V had a lower accuracy rate than VAO ($p=.01$) by 7.778 [1.432, 14.123].
There were no significant differences between the other conditions. \HDY{The results are shown in Figure~\ref{Study1-Accuracy-Rate}A.}
This result supports \textbf{H1}. This is because the accuracy rate was lowest in V and highest in VAO. }



\paragraph{\textbf{Task completion time}}



\HDY{The results are shown in Figure~\ref{Study1-Accuracy-Rate}B.}
There was a main effect on the task completion time ($F(2.362, 61.4)$ =3.084, $\eta_p^2$=.106, $p$=.045). 
\sy{Task completion time was slowest for V (M = 17.333 s [14.275, 20.392]), followed by VO (M = 16.296 s [13.415, 19.178]), VA (M = 13.889 s [11.061, 16.717]), and VAO (M = 12.926 s [10.612, 15.240]) being the fastest. 
Pairwise comparisons revealed that V was slower than VAO ($p=.027$) by 4.41 [1.498, 7.317]. 
These results support \textbf{H1}, showing that task completion time was slowest under the VAO condition.}



\paragraph{\textbf{Confidence Rating}}
\HDY{The results are shown in Figure ~\ref{Study1-Accuracy-Rate}C.}
There was a main effect on the Confidence Rating ($F(1.842, 47.889)$= 14.965, $\eta_p^2$=.365, $p$<.001). 
\sy{Confidence Rating was lowest for V (M = 5.641 [5.110, 6.172]), followed by VO (M = 5.919 [5.472, 6.365]), VA (M = 6.615 [6.448, 6.782]), and VAO (M = 6.844 [6.733, 6.955]).
Pairwise comparisons showed that V had a lower confidence rate than VA ($p=.002$) and VAO ($p<.001$). In addition, VO (M = 5.92 [5.47, 6.37]) had a lower confidence rate than VA ($p=.028$) and VAO ($p=.002$). }



These results support \textbf{H2} in the Confidence Rating, indicating that \sy{the multisensory conditions yield higher confidence ratings for responses than in the visual-only condition.}

\paragraph{\textbf{NASA-TLX}}




There was a main effect of \textbf{Performance} (F(3, 78) = 3.013, $\eta_{p}^2$ = 0.104, p = 0.035). Pairwise comparisons revealed that 
\sy{V had higher performance rate (M = 3.222 [2.452, 3.993])  than VAO (M = 2.296 [1.448, 3.145], $p=.041$).  }


A main effect of \textbf{Effort} (F(3, 78) = 4.548, $\eta_{p}^2$ = 0.149, $p = .005$) was also disclosed.  Pairwise comparisons showed that V (M = 3.519 [2.813, 4.224]) had a higher effort rate than VAO (M = 2.556 [1.762, 3.349], $p=.040$). In addition, VO  (M = 3.370 [2.693, 4.048]) had a higher effort rate than VAO ($p=.031$). These findings partially support \textbf{H3}.


\paragraph{\textbf{SSQ}}
There were no significant main effects across SSQ subscales (Nausea ($p$=0.542), Oculomotor ($p$=0.668), and Disorientation ($p$=0.488) or in participants’ perceived helpfulness ratings ($p=.059$)
among conditions. Therefore, \textbf{H4} is rejected. 








\section{Discussion}

\paragraph{\textbf{Providing multisensory cues enhances scene recognition through the portal:}} 
Our results showed that providing multisensory cues (i.e., the VAO condition) significantly improved accuracy rates, reduced task completion time, and increased participants’ confidence levels during scene recognition tasks. 
Moreover, all VO, VA, and VAO conditions yielded higher accuracy rates than the visual-only condition. 
Interestingly, although VO and VA did not significantly enhance these behavioral metrics beyond accuracy, they were associated with reductions in perceived effort, as shown in the NASA-TLX results. 
This suggests that even a single additional sensory cue can reduce cognitive load, potentially by helping participants filter and interpret ambiguous visual information. 
However, it is the combination of auditory and olfactory cues in VAO that likely provided a richer, more reliable context, reinforcing visual input and reducing perceptual uncertainty. 

Notable differences emerged in subjective ratings and participant feedback regarding the use of auditory and olfactory cues in portals. Although there was no significant difference in recognition accuracy between VA and VO, participants consistently rated audio cues as more effective and easier to distinguish than olfactory cues.
Participants reported that olfactory cues caused increased fatigue and were difficult to differentiate, whereas auditory cues were perceived as more intuitive and less cognitively demanding. 
These findings suggest that while both modalities can aid performance, auditory cues may offer a more practical and user-friendly enhancement for portal interactions in VR. 

\ic{These findings are consistent with multisensory integration theory, which suggests that the brain combines information across modalities to enhance perception and reduce uncertainty \cite{stein2008multisensory}. The VAO condition provided congruent auditory and olfactory signals that reinforced visual input, allowing participants to bind cues into a coherent representation and thus respond more quickly and accurately. 
This pattern is also compatible with cross-modal priming, where exposure to one modality (e.g., sound or scent) may activate expectations in another, supporting faster recognition and lower cognitive load~\cite{gottfried2003nose}. }

\paragraph{\textbf{Multisensory cues increase participants’ task confidence:}}
Participants reported higher confidence ratings in conditions that included additional sensory cues compared to the V condition. Notably, the VAO condition elicited the highest confidence ratings, suggesting that the combined use of visual, auditory, and olfactory cues helped participants feel more certain about their decisions.  While both VO and VA conditions also led to increased confidence compared to V, their effects were less pronounced than the full multisensory VAO condition. These findings suggest that multisensory integration, particularly when multiple senses are simultaneously engaged, can enhance users’ perceived certainty during recognition tasks in VR environments, even when visual information is limited.

\paragraph{\textbf{Multisensory cues reduce cognitive workload:}}
NASA-TLX results revealed partial support for H3. Specifically, participants in the VAO condition reported lower effort than in V and VO, and lower performance workload ratings than in V. 
These results indicate that multisensory cues can help offload some of the cognitive burden associated with interpreting limited visual information. Interestingly, participants in the V condition reported the highest self-rated performance, despite performing worst in terms of accuracy, suggesting a potential overestimation of ability when relying solely on visual information. This discrepancy highlights the value of objective metrics in evaluating performance and workload.

\paragraph{\textbf{Participants remained visually dominant despite multisensory input:}}
Qualitative feedback and post-task interviews revealed a persistent reliance on visual cues. 
In the post questionnaire, 16 participants commented they preferred VAO the most, followed by V (8), VA (3), and VO (1). 
Participants often used the visually displayed elements inside the Portal to cross-reference with audio or scent cues. 
While multisensory input aided recognition, the visual modality remained the anchor for decision-making. 
This suggests that while multisensory cues enhance performance, their role may be more supportive than primary in visually constrained tasks. 
\HDY{Their feedback was consistent with earlier research findings on the effects of multisensory cues on human perception. Previous research has shown that when visual and auditory stimuli are presented simultaneously, people tend to prioritize the visual component~\cite{colavita1974human}, known as the Colavita Visual Dominance Effect. Similarly, earlier research examining vision and olfaction stimuli together suggested that vision exerts a stronger influence on human perception than olfaction~\cite{gottfried2003nose, de2005cognitive}.
}
Future work should consider adding visual-free conditions to better isolate and understand the true effects of sensory augmentation.

\paragraph{\textbf{Simulator sickness remained unaffected by sensory cues:}}
Contrary to concerns that adding olfactory and auditory stimuli might increase discomfort, our results showed no significant differences in SSQ scores across the four conditions. This rejects H4, indicating that the introduction of multisensory stimuli did not lead to increased symptoms such as nausea, oculomotor strain, or disorientation. 
\HDY{This finding is consistent to earlier findings demonstrated the minimal effects of audio and olfaction on simulator sickness~\cite{grassini2021evaluating, zholzhanova2025virtual}. In contrast, Keshavarz et al. reported that pleasant odors may even help alleviate simulator sickness. This evidence collectively supports} the viability of incorporating scent and sound in VR environments without compromising user comfort or safety.

\subsection{Limitation and Future Work}

A primary limitation of this work lies in the constrained design of the experimental tasks. 
The scenarios used in our study may not fully capture the complexity of real-world VR portal interactions, particularly features such as dynamically resizing portals, interacting with objects within portals, or navigating directly through them. 
These simplified tasks, while useful for controlled measurement, may \sy{introduce learning effects in completing our tasks in repeated trials and} may not reflect how users engage with multisensory cues during more complex and interactive VR experiences. 
To address this, future work should explore the use of auditory and olfactory cues in more ecologically valid and dynamic contexts, such as collaborative environments, narrative-driven scenarios, or large-scale virtual spaces that demand ongoing interaction, spatial reasoning, and information retrieval. 
Moreover, while this work primarily focused on quantitative metrics, asking participants to verbally describe scenes visible through portals and their associated multisensory cues could provide deeper insight into how these cues shape comprehension of what lies beyond the portal. Such approaches would help assess the scalability and practical effectiveness of multisensory feedback in supporting immersive portal-based navigation.


A second limitation involves the olfactory feedback system. Several participants reported that lingering scents from previous trials occasionally interfered with their ability to detect new scent cues. This issue may have arisen from limitations in scent management, despite providing rest periods between trials. Because olfactory cues are more difficult to clear than visual or auditory stimuli, future work should consider more robust dispersion and removal strategies, such as active ventilation, the use of scent-neutralizing agents, or longer inter-trial intervals, when designing VR studies with olfactory stimuli.

\sy{Lastly, this study’s participant pool was constrained in diversity. In particular, the number of male participants was substantially higher than the number of female participants (21 vs. 6). 
Prior research has shown that females generally exhibit greater olfactory sensitivity and discrimination ability than males \cite{Sorokowski2019Sex}, which could have influenced how multisensory cues were perceived in our study.
Moreover, the participants were all young adults, which limits the ability to generalize our findings across other age groups. 
Olfactory sensitivity and multisensory integration are known to change with age, with declines reported in older adults and developmental differences observed in younger populations. 
These demographic constraints represent potential sources of bias. Future research should therefore recruit larger and more heterogeneous participant groups, spanning gender, age, and cultural backgrounds, to enhance the robustness and generalizability of the results.
}

\section{Conclusion}

This work investigated the effectiveness of incorporating multisensory cues—specifically auditory and olfactory stimuli—into Portal, which is a visually constrained VR interaction method.
We conducted a formal user study to evaluate how these cues influence users’ spatial understanding within VR portals. 
The findings demonstrated that the addition of multisensory cues improved participants’ ability to quickly and accurately grasp the context of virtual scenes, compared to visual-only conditions.
Importantly, while both auditory and olfactory cues contributed to enhanced accuracy, subjective responses revealed a clear preference for auditory cues. Participants reported that auditory stimuli were easier to perceive, less fatiguing, and more intuitively linked to scene content. 
In contrast, some found olfactory cues harder to distinguish and more mentally taxing, suggesting that the effectiveness of scent-based interaction may be constrained by current scent delivery technologies and scent recognizability. 
Finally, we concluded by outlining our study limitations and suggesting future research directions to explore portal-based interactions in more diverse and practical VR scenarios.
\section*{ACKNOWLEDGMENTS}
This research was supported by the National Research Foundation of Korea (NRF) grant funded by the Korea government (MSIT) (No. RS-2023-00254695).

\bibliographystyle{ACM-Reference-Format}
\bibliography{0_main}

\end{document}